\documentclass[prb, twocolumn, superscriptaddress]{revtex4-2}
\usepackage{amsmath,amssymb,bm}
\usepackage{hyperref}
\hypersetup{
    colorlinks=true,
    citecolor=blue,
    linkcolor=blue,
    filecolor=blue,      
    urlcolor=blue,
    pdftitle={Overleaf Example},
    pdfpagemode=FullScreen,
    }
\usepackage{graphicx}
\usepackage{epstopdf}
\usepackage{latexsym}
\DeclareUnicodeCharacter{2212}{-}
\usepackage[usenames, dvipsnames]{color}
\usepackage[usenames, dvipsnames]{xcolor}
\usepackage{braket}
\usepackage{float}
\usepackage[normalem]{ulem}
\usepackage{comment}
\usepackage{mathtools}
\usepackage{array}
\usepackage{tabu}
\usepackage{multirow}
\usepackage[inline]{enumitem}
\usepackage{cleveref}
\usepackage{xcolor}
\usepackage[normalem]{ulem}
\usepackage{natbib}
\usepackage{amsmath}

\begin{document}

\title{Pinning of Vortices by impurities in Unconventional superconductors}
\author{Chiranjit Mahato}
\affiliation{Department of Physics, Indian Institute of Science Education and Research (IISER) Kolkata, Mohanpur - 741246, West Bengal, India}
\author{Kun Yang}
\affiliation{Department of Physics, Florida State University, Tallahassee, Florida 32306, USA}
\affiliation{National High Magnetic Field Laboratory, Tallahassee, Florida 32310, USA}
\author {Amit Ghosal}
\affiliation{Department of Physics, Indian Institute of Science Education and Research (IISER) Kolkata, Mohanpur - 741246, West Bengal, India}
\begin{abstract}
We carry out a microscopic study of a vortex lattice in a strongly correlated, type-II, d-wave superconductor(SC) using Bogoliubov–de Gennes (BDG) formalism. In weak-coupling theory, commonly accepted truism is that a vortex binds to impurity. We demonstrate that in unconventional SCs, the binding of vortex to an impurity depends on relevant parameters. In particular, we illustrate such dependency on the sign of impurity, i.e. attractive or repulsive, as well as doping. We emphasize that this seemingly unanticipated behavior arises from strong correlation effects and is absent in weak coupling descriptions. 
\end{abstract}
\maketitle
\begin{figure*}[t]
\includegraphics[width=1\textwidth]{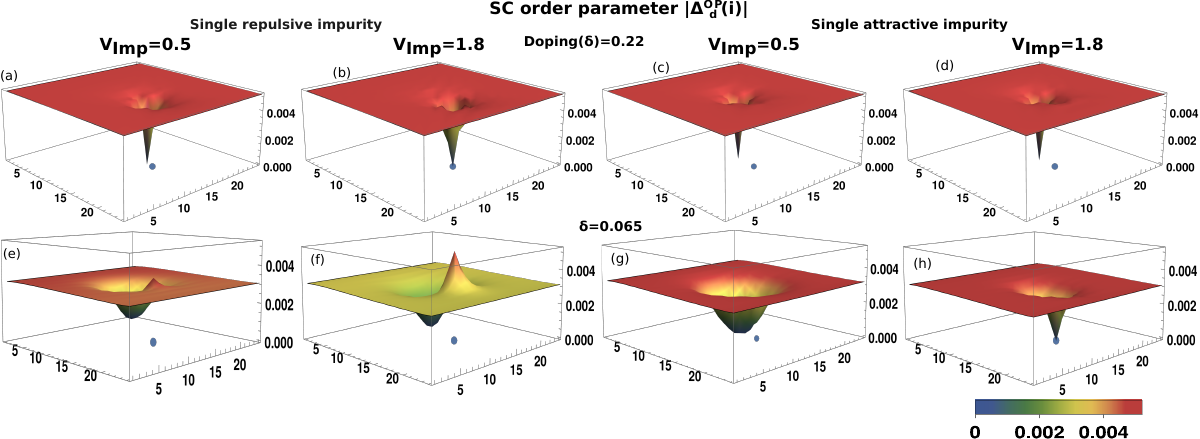}
\caption{ SC order parameter profile:Spatial profile for d-wave SC OP $|\Delta^{OP}_d(i)|$ at doping $\delta=0.22$(a,b,c,d) and $\delta=0.065$(e,f,g,h) on a magnetic unit cell of size $24\times 24$. For single repulsive impurity shows, vortex gets pinned to impurity in nearly optimally doped(a,b) case but remains unaltered in strongly underdoped(e,f) region.For a single attractive impurity vortex pinned to impurity in the strongly underdoped(g,h) case for impurity strength 1.8 or higher but not for weaker strengths and remains unaltered in nearly optimally doped(c,d) case.}
    \label{f1}
   \end{figure*}

\textit{Introduction: \textemdash }
Individual effects of disorder and orbital magnetic field on a superconductor (SC) are well established by now~\cite{RevModPhys.66.1125,RevModPhys.81.45,zhu2016bogoliubov}. However, the response of an SC to the simultaneous presence of impurities and the external magnetic field is more subtle~\cite{PhysRevB.107.L140502}, particularly for an unconventional SC~\cite{PhysRevB.69.094503}. From a technological perspective, a controlled pinning of a superconducting vortex lattice (VL) is of keen interest, because it couold enhance the critical current in a SC~\cite{Vallès2022,Haugan2004,MacManus-Driscoll2004}. While an impurity (or inhomogeneities, in a broader sense) is expected to trap a vortex in a conventional SC, unconventional SCs, often driven by strong electronic correlations, feature intriguing responses to disorder or inhomogeneity~\cite{Dam1999,Llordés2012,Gutiérrez2007}.

  The effect of disorder on both conventional and unconventional Superconductors (SCs) has been under extensive scrutiny by both theoretical~\cite{TVRamakrishnan_1989,RevModPhys.66.261,PhysRevLett.85.3922,PhysRevB.63.020505} and experimental~\cite{PhysRevLett.62.2180,Hebard,RevModPhys.79.353} techniques.
  The stark contrast in the role of non-magnetic impurities on s-wave superconductors (sSC) and d-wave superconductors (dSC) attracted initial attention, leading to a phenomenal immunity of sSC to disorder (known as Anderson's theorem~\cite{ANDERSON195926}), whereas Abrikosov and Gorkov predicted extreme sensitivity of dSC to impurities~\cite{abrikosov1960contribution,A_1_Abrikosov_1969}. Subsequent research integrated state-of-the-art experimental and theoretical techniques, comprehending predominantly the physics of disorder driven superconductor-insulator transition~\cite{Dubi2007,Gantmakher_2010,dobrosavljevic2012conductor,PhysRevLett.81.3940,PhysRevB.59.4364}.

Another route to make a pure superconductor inhomogeneous is by applying an orbital magnetic field. It is well known that the magnetic field penetrates a type-II SC by generating a periodic array of Abrikosov vortices~\cite{ABRIKOSOV1957199} which has a normal metallic core of size $\xi$, and with circulating currents around each vortex. These current rings around each vortex extend up to the scale of the penetration depth $\lambda$. With increasing $H$, the density of vortices increases and overlaps at a critical field strength $H_{c2}$, and beyond that the superconductor transitions into a metal.
Vortices are like ``punched holes" in the superconducting pairing amplitude $\Delta$. Interestingly, impurities too deplete the pairing amplitude around it. Thus, it is natural to expect, at least naively, that a type-II SC in the combined presence of impurities and orbital magnetic field, vortex centers, and impurities attract each other. Such pinning of vortices at impurities is crucial for their application in the electric power industry and even in the search of fundamental particles in the broad area of high-energy physics~\cite{Selvamanickam1998,Slimani2022}. Theoretical attempts to grasp the routes to controlled pinning have been put forward~\cite{PhysRevB.66.064509,PhysRevB.73.134515,PhysRevB.69.094503,PhysRevB.66.104502}, however, no convergence has yet been reached. 


How are inhomogeneities arising from vortices and impurity intertwined?
The pinning of a vortex by an impurity seems natural in a conventional sSC.
Both impurity and magnetic field deplete the SC pairing amplitude individually, and this costs condensation energy.
The energy cost is minimized if the depletion of pairing amplitude from the two sources (i.e. the vortex and impurity) overlaps in space. As a result, vortices (and hence the whole VL) get pinned by impurities in conventional superconductors.

The situation is more subtle for an unconventional SC. Unlike their conventional cousins, the vortices in an unconventional SC accumulate charge carriers in the core region\cite{PhysRevB.66.214502,PhysRevB.71.064504,PhysRevLett.80.3606,PhysRevLett.85.1536,hoffman2002four}. In particular, it has recently been established~\cite{PhysRevB.107.L140505,sahu2022superconducting} that the electronic density-profile in an unconventional vortex core depends on doping. 
In this backdrop, the effect of sparsely distributed impurities (effectively emulating the physics of a single impurity~\cite{PhysRevB.51.15547,PhysRevB.39.9664}) on the mixed state of a strongly correlated superconductor becomes more challenging.

In this paper, we address the issue of binding a vortex in a strongly correlated dSC in the proximity of an impurity. Our key findings are: (a) In the presence of strong correlation, the binding of vortex to impurity depends on both the nature of the impurity as well as doping. (b) The local electronic structure is mapped out. (c) We show how the LDOS at the
impurity site, which is routinely obtained from STS studies, determines the binding of a vortex to an impurity in a strongly correlated d-wave superconductor.

\textit{Model and method: \textemdash }
  We describe our two-dimensional (2D) strongly correlated d-wave superconductors (dSC) by the Hubbard model~\cite{SCALAPINO1995329}:
\begin{equation}
{\cal H}_{\rm Hubb} = -t\sum_{\langle ij\rangle,\sigma} \hat{c}_{i \sigma}^{\dagger}\hat{c}_{j \sigma}+{\rm h.c.} + U\sum_i \hat{n}_{i \uparrow}\hat{n}_{i \downarrow}-\mu \sum_{i\sigma} \hat{n}_{i\sigma}
\end{equation}
Here, $t$ and $U$ denote the hopping amplitude and on-site Hubbard repulsion respectively, $c^\dagger_{i\sigma}$ ($c_{i\sigma}$) creates (annihilates) an electron on site $i$ with spin $\sigma$ on a 2D square lattice, and $\hat{n}_{i\sigma}$ is the spin-resolved number operator.
Strong electronic correlations imply $U>\gg t$. In this limit, the low-energy physics of the above ${\cal H}_{\rm Hubb}$ is well described~\cite{K_A_Chao_1977} by the t-J model~\cite{PWAnderson_2004}. The vortices are generated by an orbital magnetic field perpendicular to the 2D-plane. Including this magnetic field, as well as impurities, our t-J model reads as:
\begin{eqnarray}
{\cal H}_{\rm t-J}&=&-t\sum_{\langle ij\rangle,\sigma}\left(e^{i \phi_{ij}} \tilde c{^\dag_{i \sigma }} \tilde c{_{j \sigma }} + h.c. \right) +\sum_{i\sigma}(V_{i}-\mu)\hat{n}_{i\sigma}  \nonumber \\ 
&+& J \sum_{\langle ij\rangle} \left ( \mathbf{\tilde{S}_{i}. \tilde{S}_{j}} -\frac{{\hat{n}_i}.{\hat{n}_i}}{4}               \right )
\end{eqnarray}
Here, the magnetic field $\mathbf{H}=\nabla \times \mathbf{A}$ and the vector potential $\mathbf{A}$ (in Landau gauge ${\bf A} = Hx\hat {y}$) is incorporated through the Peierls factor: $\phi_{ij}=\frac{\pi}{\phi_{0}}\int^j_i\mathbf{A}.d\mathbf{l}$, where $\phi_0=h/2e$ is the superconducting flux quantum, and $V_{i}$ is the impurity strength at site $i$. Note that $\tilde c^{\dagger}{_{i \sigma }}$ ($\tilde c_{i \sigma }$) appearing in the t-J model are not the standard creation (annihilation) operators, as they incorporate the restriction on double occupancy enforced by that strong onsite Hubbard repulsion. Thus, $\tilde c{_{i \sigma }}=c{_{i \sigma }}(1-n_{i\bar{\sigma}})$ is the annihilation operator on the restricted Hilbert space of no double occupancy. $\mu $ is the chemical potential which fixes the average density of the system to a desired value $\rho$. Here, ${\bf S_{i}}$ is the spin operator and the exchange interaction $J=\frac{4t^2}{U}$.

 The suppression of double occupancy arising from strong correlations are implemented via Gutzwiller approximation (GA)~\cite{PhysRevB.76.245113} in which $t$ and $J$ get adequately renormalized: $t_{ij} \rightarrow tg^{t}_{ij}$  $J_{ij} \rightarrow Jg^{J}_{ij}$, here $g^{t}_{ij}$ and  $g^{J}_{ij}$ are Gutzwiller renormalization factors (GRFs)~\cite{PhysRevB.76.245113,FCZhang_1988}. These GRFs depend on the local density as:
 \begin{equation}
 \begin{split}
 g^{t}_{ij}=g^{t}_{i}g^{t}_{j}; \hspace{0.1cm}  g^{t}_{i}=\sqrt{\frac{1-n_{i}}{1-n_{i}/2}}\\  
  g^{J}_{ij}=g^{J}_{i}g^{J}_{j}; \hspace{0.1cm}  g^{J}_{i}={\frac{1}{1-n_{i}/2}}
  \label{Eq:GFRs}
  \end{split}
 \end{equation}
 The restrictions on double occupancy is expected to reduce the effective hopping on a link, and similarly an increase of effective $J$ due to reduced double occupancy. GA is designed to emulate these effects~\cite{PhysRevB.76.245113}. Once the effects of strong correlations are incorporated using GRFs, we set up a fully self-consstent inhomogeneous mean-field theory (IMT), which is commonly termed as Bogoliubov-de Gennes (BdG) calculations~\cite{PhysRevLett.81.3940,PhysRevB.65.014501,zhu2016bogoliubov}. We refer to such Gutzwiller-augmented IMTs as GIMT calculations, whereas an IMT calculation will refer to those where all GRF's are set to unity, and hence IMT calculations will exclude the effects of strong correlations.
 
In our calculations, we consider systems with $16 \times 8$ (magnetic) unit-cells where each unit-cell is of size $24 \times 48$ (lattice spacing is taken as unity). We chose a field-strength which corresponds to two superconducting flux-quanta through each unit-cell. The resultant vortices constitute a square vortex lattice. 
\begin{figure*}[!htb]
    \includegraphics[width=1\textwidth]{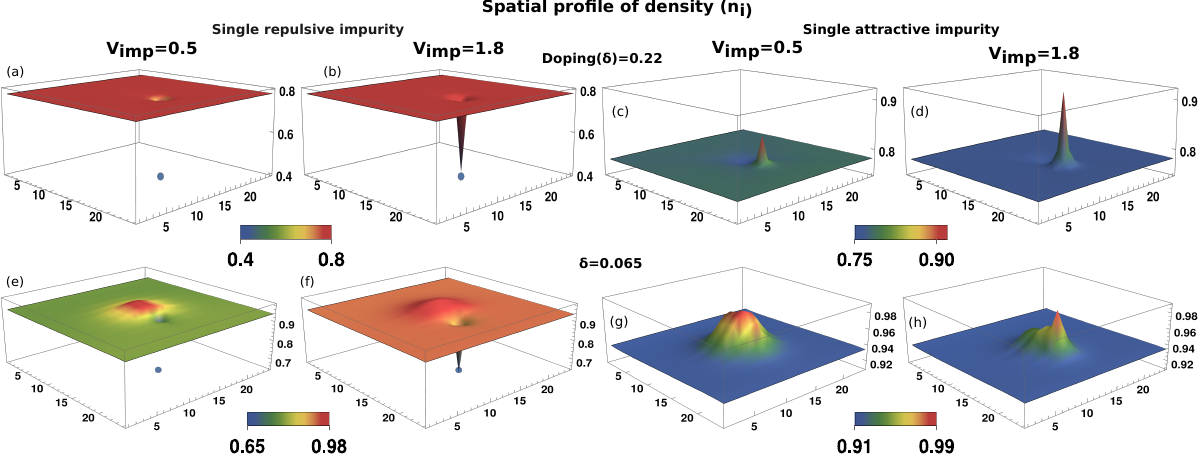}
   \caption{Local charge density: Spatial profile of local charge density around a vortex core and at impurity site for different doping $\delta$. With $\delta=0.22$, there is a dip at both the vortex core and impurity site for a single repulsive impurity, whereas, for doping $\delta=0.065$ charge accumulates around the vortex core in contrast to the dip at the impurity site. For a single attractive impurity with $\delta=0.22$, there is charge accumulation at the impurity site in contrast to the dip around the vortex core. Whereas for doping $\delta=0.065$ charge accumulates around the vortex core and at the impurity site in case of attractive impurity. } 
    \label{f2}
   \end{figure*}
In our simulation, we consider a square vortex lattice instead of energetically favorable triangular vortex lattice in continuum. For our simulation on limited system sizes, a square vortex lattice is favorable as it is consistent with the symmetry of the underlying microscopic lattice under consideration. We start our BdG simulation with order parameter taken from the Abrikosov solution of the Ginzburg-Landau equations in the presence of the magnetic field.
   
Here we choose our rectangular magnetic unit cell ($L \times 2L$) containing two flux quantum in such a way to place the center of each vortex at the center of the half of magnetic unit cell of size $L \times L$.

We then place a single impurity on a lattice site three lattice spacing away from the center of a vortex (i.e. the position of the vortex-center in the pure system ). We then solve the above $H_{t-J}$ fully self-consistently using the scheme of GIMT~\cite{Chakraborty_2014, PhysRevB.63.020505}. The goal is to understand how the new vortex position is influenced upon including the impurity. We present all energies in units of $t$ and we consider $U=12t$ resulting into $J=\frac{4t^2}{U}=0.33$~\cite{PhysRevB.52.615}. Our GIMT results are also contrasted with IMT calculation, to highlight the role of strong correlations in determining the interplay of vortices and impurities.


\textit{Results: \textemdash }
We start presenting our results on dSC pairing amplitude: $\langle\tilde{c_{i\sigma}}\tilde{c}_{j\bar{\sigma}}\rangle_{\psi} \approx g^{t}\Delta_{ij}$~\cite{PhysRevB.78.115105,PhysRevB.96.134518}. Here $\langle \cdots \rangle_{\psi}$ denotes the expectation value, evaluated using GIMT formalism, and $\Delta_{ij}=\langle c_{i\sigma} c_{j\bar{\sigma}}\rangle_0$ -- now the expectation value $\langle \cdots \rangle_0$ is calculated utilizing the IMT framework. 


\textit{d-wave SC order.}
The spatial profile of the local dSC order parameter :
\begin{eqnarray}
\Delta^{OP}_d(i) &=& \frac{J}{4} \left\vert [g^{t}_{i,i+\hat{x}}\Delta^{\hat{x}}(i)+g^{t}_{i,i-\hat{x}}\Delta^{\hat{-x}}(i)\right. \nonumber \\ 
&-& \left. e^{ibx}g^{t}_{i,i+\hat{y}}\Delta^{\hat{y}}(i)-e^{-ibx}g^{t}_{i,i-\hat{y}}\Delta^{\hat{-y}}(i)]\right\vert
\label{Eq:DelOP}
\end{eqnarray}
(here $b=\frac{H}{\phi_0}$) is shown on half unit cell containing one SC flux quantum in Fig.~\ref{f1} for different parameters. Fig.~\ref{f1}(a) presents pairing amplitude for near optimal doping ($\rho=0.78$) when a repulsive impurity (of strength $V_0=0.5$) is attached to a site which is $3$ lattice spacing away from the vortex position of a clean square vortex lattice. The resulting self-consistent pairing amplitude indicates that the vortex tend to follow the impurity.
Fig.~\ref{f1}(b) shows results similar to panel (a) but for stronger impurity strength, $V_0=1.8$. In this case the vortex center moves to the impurity location. Fig.~\ref{f1}(c,d) show results similar to Fig.~\ref{f1}(a,b) respectively, but for attractive impurities ($V_0=-0.5$, $-1.8$ respectively). The results demonstrate that the vortex does not move to bind to the attractive impurities, i.e. the vortex lattice remain unaltered upon introducing the impurity.

Fig.~\ref{f1}(e,f) depicts vortex profile in the presence of repulsive impurities similar to panels (a,b) but for strongly underdoped case ($\rho=0.935$). Self-consistent pairing amplitude does not move the vortex centres to the impurity site. In fact, the results show a seemingly unusual enhancement of dSC pairing amplitude at the impurity site! Also, this enhancement gets sharper with the increase of the impurity strength.

In order to analyze this unanticipated enhanceent of $\Delta_{i_{0}j}=g^{t}{(i_{0},j)}\Delta^{0}_{i_{0}j}$, where $i_{0}$ is the impurity site and $j$'s  are its nearest neighbours, we have studied the spatial profile of $g^{t}{(i,j)}$ and $\Delta^{0}_{ij}$, which confirms that this elevation is due to the enhanced GRF ($g^{t}_{i_{0}j}$) connecting to the impurity site. This occurs because the local density $n_{i_{0}}$ is depleted due to the presence of the impurity, causing a local rise in $g^{t}{(i_{0},j)}$ as per Eq. (\ref{Eq:GFRs}). $\Delta^{0}_{i_{0}j}$ -- the pairing amplitude evaluated at the unprojected space, on the other hand undergoes a weak depletion at $i_{0}$, yet the local enhancement of $g^{t}_{i_{0}j}$ overpowers that depletion, and controls the profile of $\Delta_{ij}$. A detailed analysis is provided in Appendix C.

Results for a single attractive impurity in the strongly underdoped region is shown in Fig.~\ref{f1}(g,h). Interestingly, the binding of vortex to the impurity depends on its strength in this case. For example, the vortex center moved to the impurity position for stronger impurities e.g. for $V_0=-1.8$ or higher, but not for weaker attractive impurity.

Thus, we see that unlike conventional superconductors, where a vortex always binds to an impurity, here things depend crucially on various parameters for a strongly correlated dSC. In the following we proceed to develop a comprehensive understanding of this behavior.  

It is interesting to note that the binding of a vortex to an impurity also controls the shape of the vortex in an interesting manner in the underdoped region, as described below. It has been recently argued that the vortex of a strongly correlated dSC takes the shape of a flat bottom bowl in the strongly underdoping region~\cite{PhysRevB.107.L140505}. We also witness the same for cases when vortex does not bind to impurities. On the other hand, it assumes the usual conical shape whenever the vortex binds to an impurity, irrespective of its sign.

In case of results from IMT calculations, pinning of vortex to the impurity does not depend on the nature and strength of the impurity. Here vortex is always pinned to impurity to minimize the energy cost associated with bending the superconducting order parameter. This is illustrated further in the supplementary materials (SM).



\textit{Local charge density at the vortex core.} We have shown in Fig.~\ref{f2} the local charge density ($n_i$) in the vortex core region  and around the impurity site (for both repulsive and attractive impurities). These results are for dopings $\delta=0.22$ and $\delta=0.065$ and are displayed on half of the magnetic unit cell of size $24\times24$ which encloses a single vortex.

Along with the dSC pairing amplitude, local charge density at the vortex core, as well as that at the impurity site play a crucial role in determining the pinning of a vortex to the impurity, as we discuss below. A clean system (without any impurity) near optimal doping ($\delta=0.22$) features a weak dip in local density at the vortex core, whereas charge density accumulates in the vortex core as $\delta$ is progressively reduced. This charge accumulation reaches close to unity (the maximum possible value in GIMT calculations) at the vortex center for strong underdoping ($\delta=0.065$). Such enhancement at underdoping originates from the close proximity of the system to a Mott insulator at the half-filling. As a result, the nature of the normal state at a clean vortex core (i.e. the local phase upon depletion of pairing in the vortex core) changes from metallic to Mott insulating as we march towards strong underdoping~\cite{PhysRevB.107.L140505}.

This change in the nature of vortex core upon doping is only realized in GIMT calculations, which highlights the role of strong correlations on the charge distribution in the vortex core. An IMT calculation, on the other hand, always shows a dip in local charge density at the vortex core irrespective of the doping level. Thus, in order to reduce the energy cost, pinning of the vortex depends only on the profile of the dSC pairing amplitude and is independent of doping level, or the sign of impurity.

On the other hand, an impurity in a homogeneous dSC (i.e. without orbital field, and hence vortices) modifies the charge density profile depending on its sign (i.e. the impurity being attractive or repulsive). An attractive impurity naturally accumulates charge carriers (until the maximal value of unity) prohibiting double occupancy arising from strong correlations. A repulsive impurity, by itself, suppresses local electronic density below its average value. Thus, whether a vortex binds to a nearby impurity in case of a strongly correlated dSC is dictated by the "compatibility" of the local density profile arising out of the two, which in turn depends on various factors, e.g. the doping level, the sign of impurity and its strength, as we illustrate below.


\begin{figure}[t]
    \includegraphics[width=0.5\textwidth]{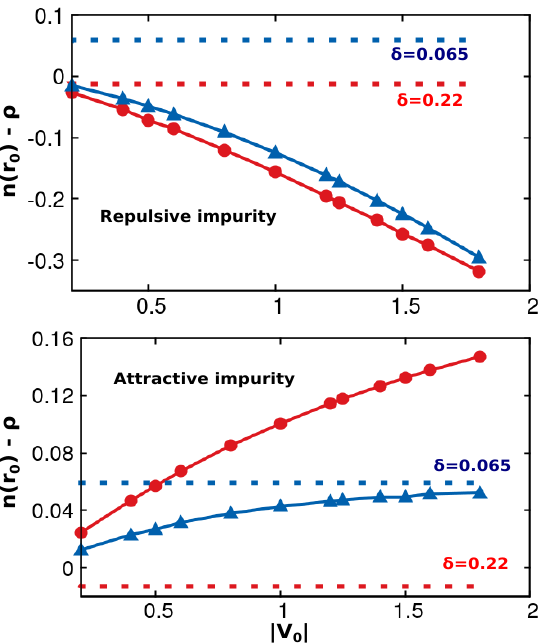}
\caption{Response of local charge density:This figure shows $n(r_0)-\rho$ profile for a clean vortex without impurity (dotted line) and a single impurity without vortex with varying impurity strength (solid lines connecting dots). $r_0$ stands for vortex core and impurity sites respectively. 
For $\delta=0.22$, there is always a dip at vortex core(red dot), and for $\delta=0.065$, the charge accumulates at the core(blue dot). $n_{core}-n$ negative for repulsive impurity and positive for attractive impurity. Its value increases with the increase of impurity strength for both repulsive and attractive impurities. } 
    \label{f3}
   \end{figure}
   \begin{figure*}[!t]
\includegraphics[width=1\textwidth]{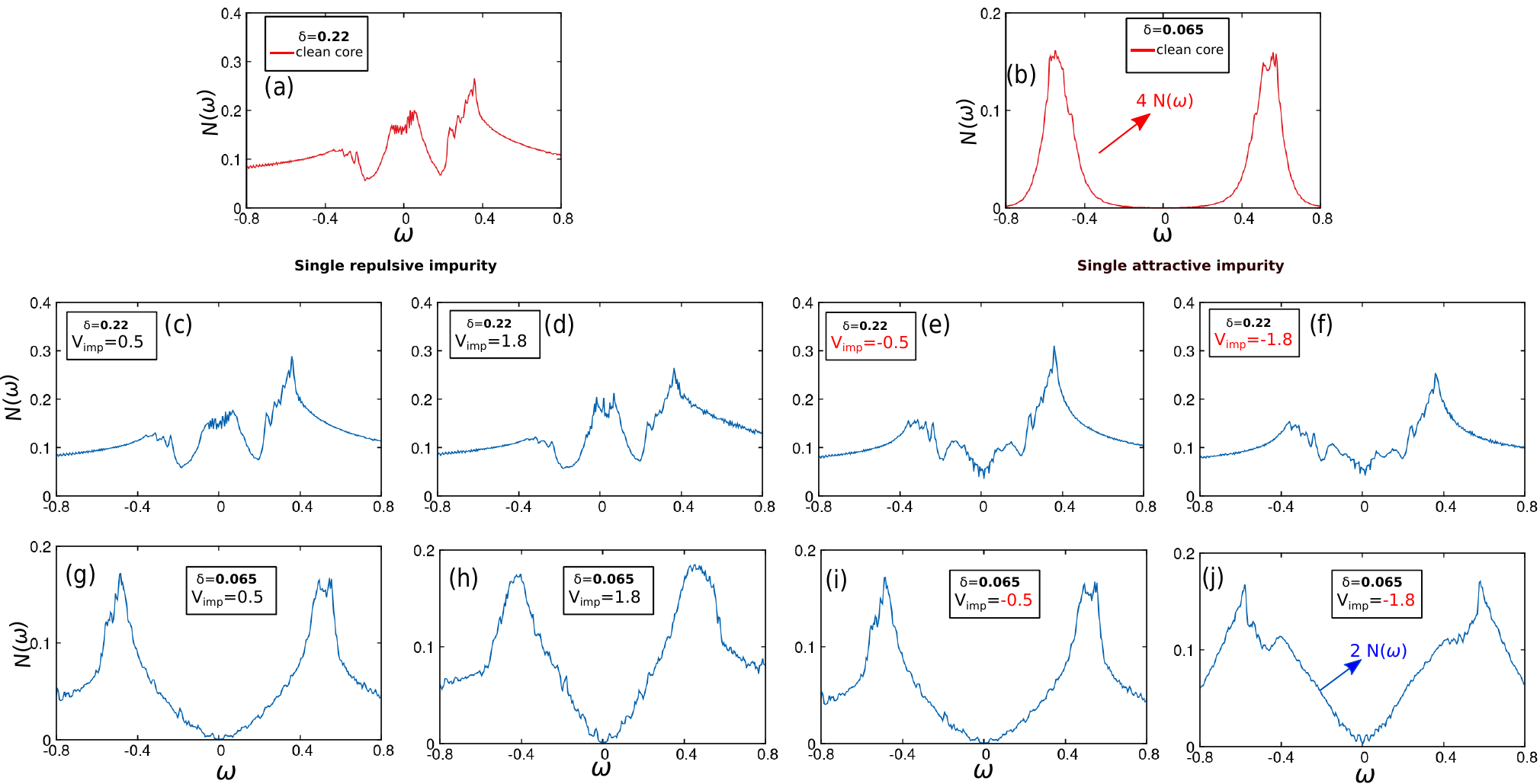}
\caption{Local density of states: LDOS at clean vortex core for $\delta=0.22$(a),$\delta=0.065$(b), and at the impurity site (for both repulsive and attractive) for the same two doping values. For repulsive impurity(c,d), in the near optimally doped region LDOS similar to that of the clean vortex core(a) but differs for a single attractive impurity. In strongly underdoped case, the LDOS  profile does not match that of the clean vortex core(b). Though for the attractive impurity of strength 1.8, vortex pinned to impurity but due to the change of the nature of SC OP profile, LDOS at the impurity site do not mimic the LDOS at the clean core. For clarity, the vortex core LDOS in panels (b and j ) has been scaled up by a factor of 4 and 2, respectively.  } 
\label{f4}
   \end{figure*}
\textit{Response of local charge density due to simultaneous presence of vortex and impurity.}
Riding on the insights developed above, we look into the local density profiles in two different situations separately: In one, with only the perfect periodic vortex lattice in the clean strongly correlated dSC, and the other with a single impurity (cases of repulsive and attractive ones), whose strength is being tuned. The competing (or collaborating) effects of the vortex and impurity individually on the response of local density is presented in Fig.~\ref{f3}. Here, a dotted trace for the departure of the density at the vortex center from its average value, $\delta n=n(r_0) -\rho$, represents the results when only the vortex lattice is present (i.e. no impurity) and the solid ones are the results for impurities with different strengths in the absence of any orbital fields. For the case of vortices alone, we notice a charge depletion for $\delta=0.22$ (near optimal doping), and a charge accumulation in the  strongly underdoped case ($\delta=0.065$).
 Since these are results in the absence of impurities, they show up as a straight lines as a function of the magnitude of the impurity strength $\vert V_0 \vert$.

Next we address the dependence of $\delta n$ on $\vert V_0 \vert $, when our system supports a single impurity in the absence of vortex lattice (i.e. $H=0$). A single repulsive impurity produces a dip in $\delta n$ irrespective of doping, and the depth of this dip increases with the enhancement of the impurity strength. Due to the compatibility of the local charge density profile (i.e. a dip in $\delta n$ both due to the vortex lattice, as well from the impurity) in case of a repulsive impurity and at optimally doped region, a vortex gets pinned to the impurity. By the same token, it does not bind to the impurity when the doping level is set to strong underdoping, because the compatibility of the local density profile from the two independent resources is lost. Similarly, for attractive impurity, there is charge accumulation at the core for both near optimally doped and strongly underdoped cases. Now, whenever there is a match of accumulated charge density due to both vortex and impurity, vortex gets pinned to impurity in the strongly underdoped region. This matching of the density profile depends on the strength $\vert V_0 \vert$ of the attractive impurity. As mentioned above, the results reflect the same: vortex gets pinned to an attractive impurity in the strongly underdoped case for an impurity strength of $V_0=-1.8$ but the structure of the vortex lattice remains unaltered for a weaker impurity $V_0=0.5$. In contrast, for attractive impurities in the near optimally doped region vortex does not prefer charge accumulation; therefore vortex does not pin to impurity.


In a sharp contrast, the IMT results show no effects of local density profile on the binding of a vortex to the impurity.\\
   \textit{LDOS at clean vortex core and impurity site.}Pinning of vortex to impurity in conventional superconductors is largely independent of doping or the sign of the impurity.
   As discussed, pinning of a vortex to an impurity requires a compatibility of the reorganization of the local electronic density arising from the two sources -- the vortex and the impurity. As a result, the LDOS features the compatibility as well. Thus the presence of the impurity (in this case of pinning of vortices at impurities) does not alter the qualitative features of the LDOS at the vortex core.
   This, however, is not true when the vortex does not bind to the impurity for certain range of the parameter space, as we already discussed. Naturally, the LDOS at the impurity site under this situation will differ significantly from that at the clean vortex core.
Within GIMT calculations, we evaluate LDOS using:
\begin{eqnarray}
N(i,\omega)&=&\frac{1}{N}\sum_{k,n}g^t_{ii}[|u^k_{n}(i)|^2\delta(\omega - E_{k,n})  \nonumber \\
&+&|v^k_{n}(i)|^2\delta(\omega + E_{k,n})].
\end{eqnarray}
Here, $N$ is the system-size, and $\left\{ u^k_{n}(i),v^k_{n}(i) \right\}$ are the Bogoliubov wave functions, $\left\{E_{k,n}\right\}$ corresponding to energy eigenstates.  
 The behavior of the calculated LDOS satisfy the aforementioned expectations, and is demonstrated in Fig~\ref{f4} for doping $\delta=0.22$ and $\delta=0.065$, where panel (a) and (b) shows LDOS profile at a clean vortex core. In this case, LDOS exhibit a low-energy core states (LECS) 
for doping $\delta=0.22$(Fig.~\ref{f4}(a)) and  a hard U-shaped gap for doping $\delta=0.065$(Fig.~\ref{f4}(b))~\cite{PhysRevB.107.L140505}.
On the other hand, Fig.~\ref{f4}(c-j) represent the LDOS at the impurity sites for different sets of parameters ($\delta$, and $V_0$).
In case of panel (c),(d) representing ($\delta$,$V_0$)=(0.22,0.5),(0.22,1.8), vortices bind to impurities, as explained in the previous section. Hence, these LDOS profiles have a significant quantitative match with the two in panels (a) and (b). On the otherhand, LDOS in panel (e),(f) and (g),(h),(i) for($\delta$,$V_0$)=(0.22,-0.5),(0.22,-1.8) and (0.065,0.5),(0.065,1.8),(0.065,-0.5), do not match with those panel (a) and (b), because impurity does not pin to vortex.
An interesting departure from the above picture occurs at strong underdoping ($\delta=0.065$), where the proximity to Mottness in GIMT calculations changes the shape of the vortices to flat-bottom bowls from the usual conical ones~\cite{PhysRevB.107.L140505}. On the other hand, binding of an attractive impurity ($V_{0}=1.8$) restores the usual conical shape of the vortex, affecting the local Mottness. As a result, the U-shaped LDOS is lost yielding a standard d-wave nature of the low lying LDOS.

Therefore, it is possible to determine if the vortex core becomes pinned to the impurity by examining not only the dSC order parameter profile but also the local density of states at the impurity site -- an experiment routinely performed using scanning tunneling spectroscopy (STS)~\cite{RevModPhys.79.353}.

{\textit{Conclusion.}} We reported how SC OP profile and local charge density play an important role in determining the binding of vortex to impurity for unconventional SCs in contrast to conventional SCs. In unconventional SCs, the nature of vortex core changes from metallic to Mott insulator. Hence in the GIMT  case vortex gets pinned to the impurity whenever there is a match of the local charge density profile between vortex core and the impurity site in addition to the order parameter profile. We verified this by not only looking at the spatial profile of dSC order parameter but also at LDOS at the impurity site.\\

\textit{Acknowledgement.} All authors acknowledge support from the Scheme of Promotion of Academic and Research Collaboration (SPARC) Grant No. 460, MoE, Govt.~of India.~KY's work is supported by National Science Foundation Grants No. DMR-1932796 and DMR-2315954, and performed National High Magnetic Field Laboratory, which is supported by National Science Foundation Cooperative Agreement No. DMR-2128556, and the State of Florida.\\




\appendix
\renewcommand\appendixname{Appendix}

\begin{figure*}[t]
		\centering
			\centering
			\includegraphics[width=1\linewidth]{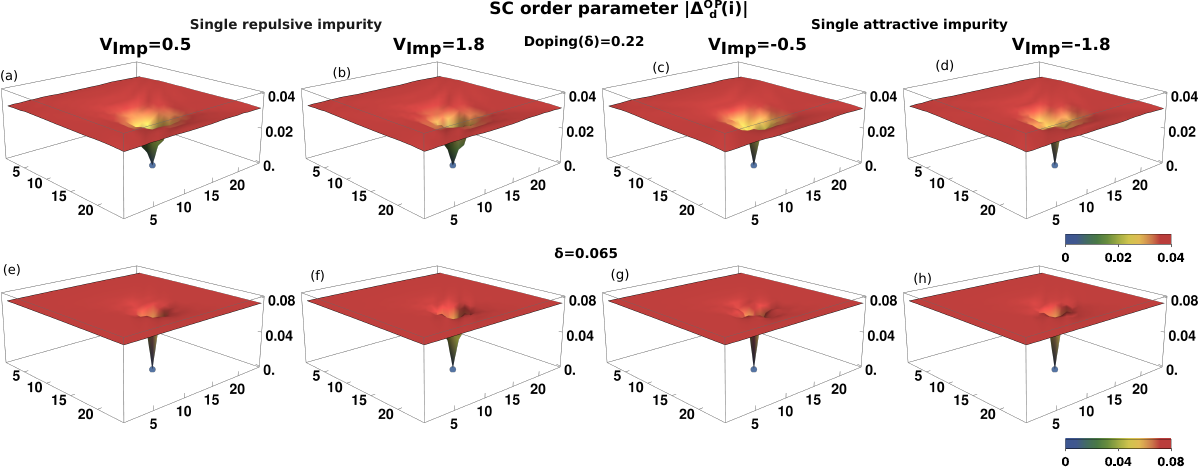}
			\caption{ Superconducting order parameter profile: Spatial profile for d-wave SC OPA $|\Delta^{OP}_d(i)|$  for repulsive impurity at doping $\delta=0.22$(a,b) and $\delta=0.065$(e,f) and attractive impurity at doping $\delta=0.22$(c,d) and $\delta=0.065$(g,h)  on a magnetic unit cell of size $24\times 24$ for IMT calculations. Irrespective of doping and the nature of impurity, vortex binds to impurity.}
    \label{f5}
  \end{figure*}

\begin{figure*}[t]
    \includegraphics[width=1\textwidth]{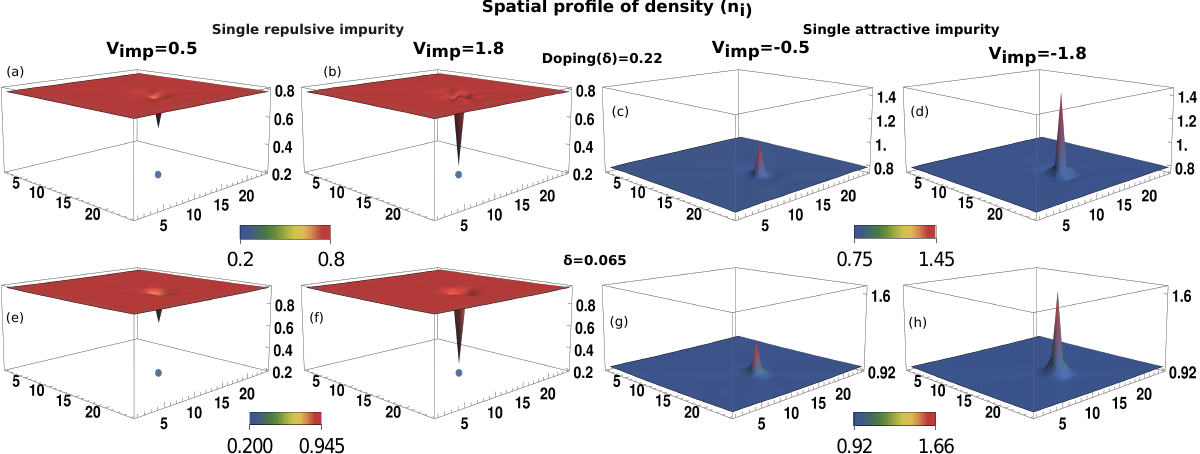}
\caption{ Local charge density: Spatial profile of local charge density around a vortex core and impurity for different doping  $\delta$(0.22,0.065) within IMT calculation.  } 
    \label{f6}
   \end{figure*}

\centerline{\textbf{Appendix}}
In order to adequately emphasize the role of Gurzwiller correlations in our findings in the main text, we provide in the appendices below the results from the IMT calculations (see main text). These IMT results ignore the effects of strong correlations by setting all Gutzwiller factors(GRFs) to unity. In the following sections, we wish to emphasize the stark differences between the GIMT and IMT solutions. The details of the formalism of IMT and GIMT calculations for a d-wave vortex lattice can be found in the supplementary information of Ref.~\cite{PhysRevB.107.L140505}.

\section{Order parameters and Local density of states(LDOS):}
\label{ORDER PARAMETERS(OP) AND LOCAL DENSITY OF STATES(LDOS):}
 Below we discuss the spatial profile of d-wave superconducting(dSC) Order parameter amplitude(OPA), local charge density, and LDOS for IMT calculations. In IMT, OPA is equivalent to pairing amplitude.
 The stark contrast of these results with those from GIMT calculations presented in the main text is evident. This demonstrates that strong correlations play a crucial role in determining the underlying physics.

{\bf dSC order parameter:} Fig.~\ref{f5} shows the spatial profile of dSC OPA in a unit cell containing one vortex of the vortex lattice(VL), plus an impurity indicated by a dot, for both attractive and repulsive impurity, as well as for two doping levels, $\delta=0.22,0.065$. These results, when contrasted with the GIMT finding in Fig.~\ref{f1} from the main text, indicate that within the IMT framework:\\
(a) An impurity pins a vortex, irrespective of the sign of impurity or doping level.\\
(b) Vortices maintain conical shape even at strong underdoping, unlike in GIMT results where their shape changes to a "flat-bottom bowl".

{\bf Local charge density:}
 We have presented the spatial profile of local charge density within IMT calculation in Fig.~\ref{f6}. There is a small charge dip at the vortex core for both doping  $\delta=0.22,0.065$. On the other hand, in these two extreme values of doping, there is a charge dip at the repulsive impurity site and charge accumulation at the impurity site for attractive impurity.

\begin{figure*}[t]
		\centering
			\centering
		\includegraphics[width=1\linewidth]{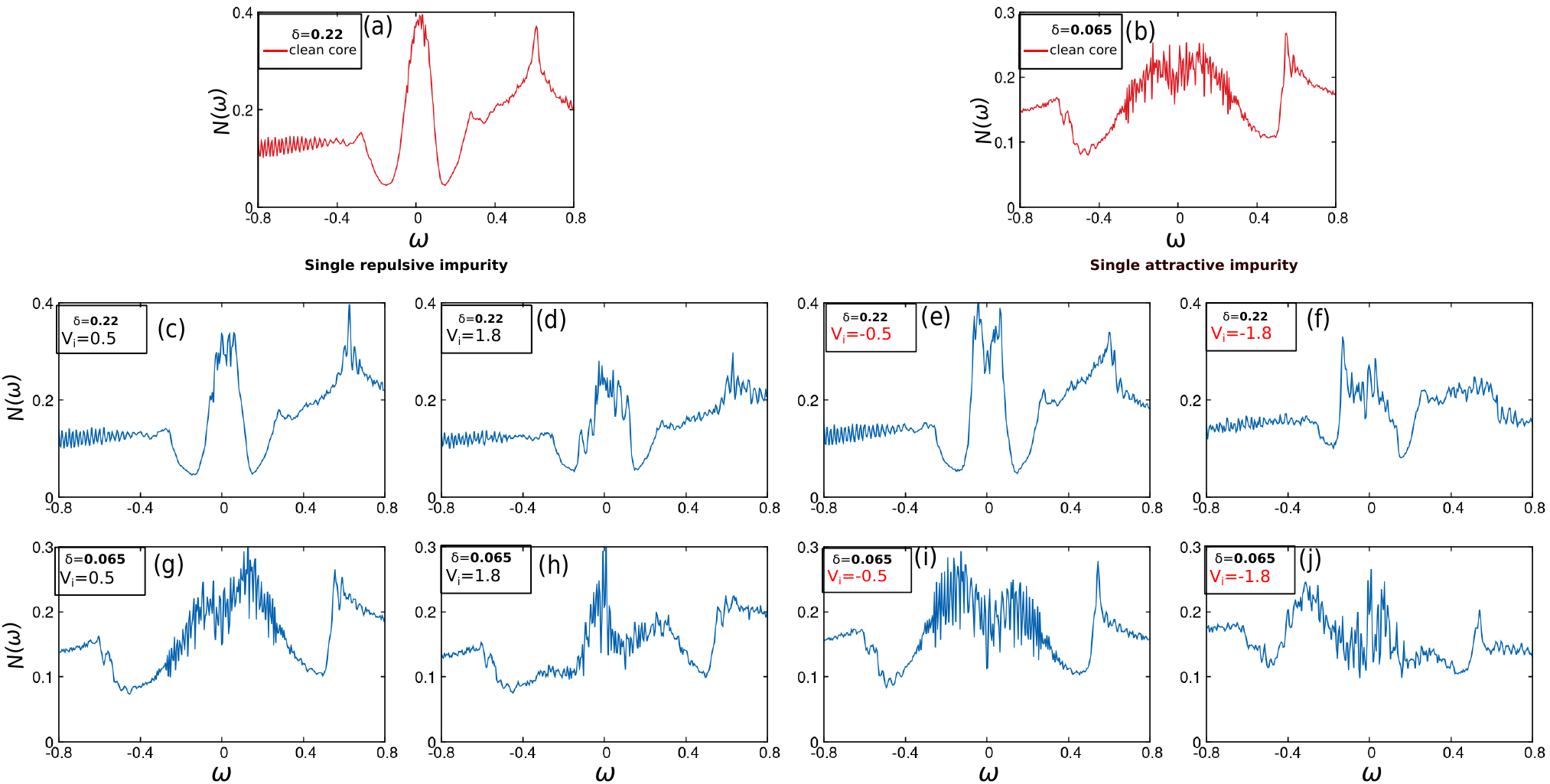}
			\caption{ Local density of states: LDOS at clean vortex core for $\delta=0.22$(a), $\delta=0.065$(b), and at the impurity site (for both repulsive and attractive) for the same two doping values within IMT calculation. Independent of doping and the nature of impurity, LDOS at the vortex core and impurity site shows the the presence of zero-energy core states.}
 \label{f7}
 \end{figure*}

{\bf Local density of states.} LDOS at clean vortex core and impurity site for $\delta$=0.22 and  $\delta$=0.065 within IMT calculation is presented in Fig.~\ref{f7}. LDOS at the clean vortex vortex core and at the impurity site carry zero-energy core states irrespective of the nature of impurity and doping. These should be contrasted with the GIMT results shown in Fig.~\ref{f4}. of the main text.
 
 These demonstrate the sharp contrast between the IMT and GIMT results, signifying the role of strong correlation in determining the binding of a vortex to impurity.

 \begin{figure*}[t]
		\centering
			\centering
		\includegraphics[width=1\linewidth]{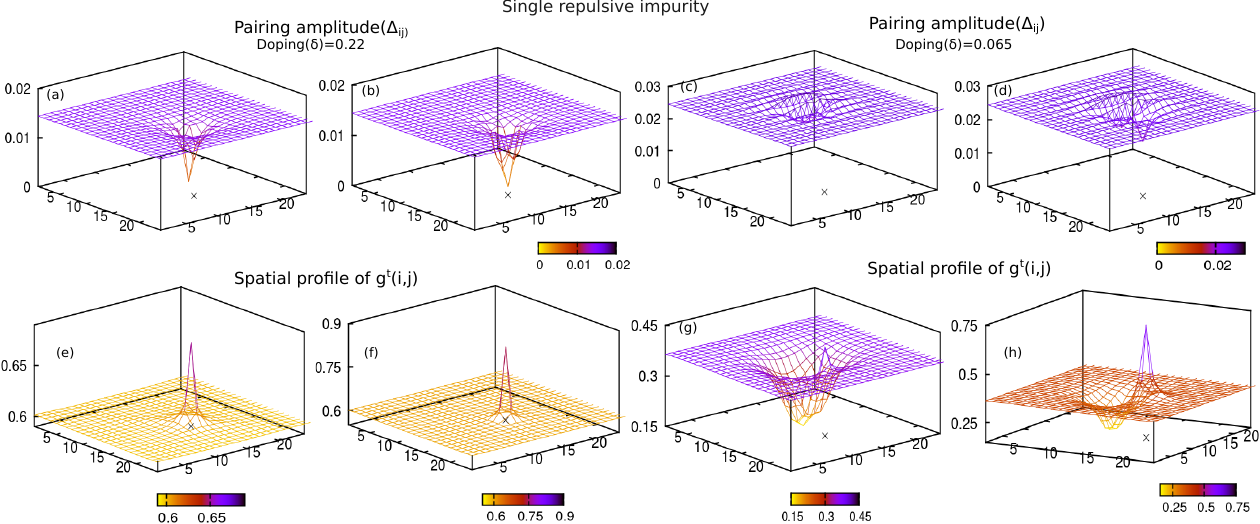}
			\caption{Spatial profile of pairing amplitude($\Delta_{ij}$) and GRF($g^{t}(i,j)$): Spatial profile of pair amplitude and GRF  for doping($\delta$)=0.22(a,b) $\And$(e,f)  and $\delta$=0.065(c,d) $\And$ (g,h) is shown for half of the magnetic unit cell. }
 \label{f8}
 \end{figure*}

\section{Intriguing peak in dSC order parameter amplitude on a repulsive impurity location at strong underdoping.}
\label{Comprehension of the intriguing feature in the order parameter shown in FIG. 1.(e,f) of the main text:}
The spatial profiles of dSC OP inFig.~\ref{f1}(e,f) show an intriguing peak. Here we present the comprehension of such a feature. The spatial profile of the dSC OP is determined by the following two factors:\\
(i) The spatial structure of Gutzwiller factor(GRF), $g^{t}(i,j)$.\\
(ii) The spatial profile of the pairing amplitude ($\Delta_{ij}$, i.e. the dSC OP without the Gutzwiller factors).\\
In the presence of an impurity, spatial profiles of both $g^{t}(i,j)$ and $\Delta_{ij}$ get modified. Interestingly, the two modifications act against each other in the case of a repulsive impurity and at underdoping, as shown in Fig.~\ref{f8}. 
The resultant dSC order parameter profile, capturing both these contributions generates a spatial pattern as shown in Fig.~\ref{f1}(e,f) in the main text.
 In the case of overdoping, the vortex binds to the impurity, and the pairing amplitude $\Delta_{ij}$ vanishes at the impurity site. As a result, such a feature is absent (See Fig.~\ref{f1}(a,b) in the main text).

\bibliography{ref}

\end{document}